\documentstyle[preprint,aps]{revtex}
\begin{document}
\draft
\title{Concerted motion of protons in hydrogen bonds of DNA-type molecules.}

\author{V.L.Golo $^{1, 3}$\, , E. I. Kats $^{2, 4} $\,, M.Peyrard  $^{3}$}

\address{$1$ Department of Mechanics and Mathematics \\
Moscow University, 119899 Moscow, Russia ; \\
$^2$ Laue-Langevin Institute, F-38042, Grenoble, France ; \\
$^3$ Lab. Phys. ENS Lyon, 69364, Lyon, Cedex 07, France; \\
$^4$ L. D. Landau Institute for Theoretical Physics, RAS \\
117940, Kosygina 2, Moscow, Russia.}
\date{\today}

\maketitle

\begin{abstract}
We study the dynamical behaviour of the proton transfer in the hydrogen bonds
in the base-pairs of the double helices of the DNA type. Under the
assumption that the elastic and the tunnelling degrees of freedom may
be coupled, we derive a non-linear and non-local Schrodinger equation (SNLNL)
that describes the concerted motion of the proton tunnelling.
Rough estimates of the solutions to the SNLNL show an intimate interplay
between the concerted tunnelling of protons and the symmetry of double helix.
\end{abstract}


I.

The recent direct observation of coherent proton tunnelling in macromolecules
\cite{1} has focused the attention on systems allowing for the nontrivial
dynamics of protons contained in hydrogen bonds. Perhaps, the most significant 
system of the kind is the DNA molecule.
The problem of the proton transfer inside the hydrogen bonds of the
double helix representing a molecule of DNA, has a long and rich
history. The importance of the phenomenon was noticed soon after the DNA
double helix had been discovered (see \cite{SA}).
In fact from the chemical point of view the proton transfer is a so-called
tautomerization reaction, that is a kind of transition that preserves the constituent
atoms of a compound but at the same time changes their mutual positions.
It has been believed that the tautomerism transition could provide a mechanism
for genetic mutations \cite{SA}.

Later
Crick \cite{CR} suggested that the mutations could be due to 
conformational changes within the double helix, the so-called
"wobbling" \cite{SA}. By now there has been no definite conclusion as to
whether the wobbling or the tautomerism are responsible for the occurrence
of mutations. The recent work on the i-motif of DNA, in which the
tautomerism, rare in the usual DNA, is the rule, and not an exception
\cite{GL} has given a new impetus to the problem of proton transfer.

The phenomenon of tautomerism in the base pairs of DNA has been studied
extensively, both experimentally and theoretically \cite{SA}, 
\cite{CI} , \cite{Dre},
but for
individual single molecules of purines and pyrimidines. Here it should be
noted that even though one has got a considerable body of information
about the chemical reaction corresponding to the tautomer transition,
such as its constant of reaction, and even managed to find an estimate for
the concentration of tautomeric forms, $10^{-4} - 10^{-5} $ mol/l,
for the B-type of DNA, there is a little knowledge as far as the dynamics
of the proton transfer accompanying it in DNA is concerned. In fact, the
estimates for the transition frequency widely diverge, and generally 
it is believed to be within $10^{6} - 10^{11} \, s^{-1} $, \cite{GL}.

In our opinion the unsatisfactory state of the art as regards to the proton
transfer, and the tautomeric dynamics, in the DNA, is to a certain extent
due to the absence of theoretical models of the phenomenon. In the present
paper we would like to draw attention to the fact that {\it if the tautomeric
transitions are coupled with a change in elastic properties of the DNA
molecule, one could expect a concerted tunnelling of the protons in the
hydrogen bonds}. Taking into account the extremely sophisticated nature of
the system, i.e. the DNA molecule, we aim at studying it in a 
qualitative framework, which takes into account, the main features of
the molecule, i.e. the presence of
{\it the two strands, the helical structure, or the winding
symmetry, and the dynamics of protons} considered within the approximation
of two-level systems.

II.

Aiming at a simple model, we shall describe the states 
of the hydrogen bonds as those of a one-dimensional Bose oscillator,
the two-level requirement being accommodated by considering only 
its ground state
and the lowest excited one. In this sense the tautomeric
reaction of the proton transfer is described by the excited state of
an oscillator.
Thus, we shall consider the protons inside the hydrogen bonds of a molecule of
DNA as a quantum system described by the Bose operators $b_n^+ , b_n $
subject to the usual commutation relation
$$
[b_n^+, b_m] = \delta _{n m}
$$
in which $n, m$ are the indices of the corresponding sites of hydrogen
bonds. We shall suppose that the system of the protons is in a weakly
excited state that can be cast in the form of wave function
\begin{eqnarray}
\label{b111}
|D(t)> = \sum _n A_n(t) b_n^+ | 0> 
\end{eqnarray}
in which $A_n(t)$ are complex amplitudes subject to the relation
$$
\sum _{n}|A_n|^2 = 1
$$
The Hamiltonian of the total system, that is the protons and the elastic
part corresponding to the sugar-phosphates, reads
\begin{eqnarray}
\label{a1}
H = H_H + H_Y + H_I
\end{eqnarray}
where $H_H$ is the proton Hamiltonian
\begin{eqnarray}
\label{a11}
H_H = \sum _{n} E_0 b_n^+ b_n 
- \kappa \sum _{n}(b_{n+1}^+ b_n + b_n^+ b_{n+1}) \, , 
\end{eqnarray}
where $E_0$ is the level splitting between two states of the proton in a hydrogen bond, and $\kappa $ is
the tunnelling probability.
$H_Y$ is that of the elastic part, and $H_I$ is the interaction between
them. We shall consider the elastic part of the system as the classical
one, and even neglect its kinetic energy. The reason is that
the characteristic times for the elastic vibrations of the DNA molecule
are usually estimated as being in the region of $10^{-11} - 10^{-13} $s,
whereas the proton tunnelling is alleged to be within $10 ^{-6} - 10^{-11}$
s (more exact figures are unavailable; according to \cite{CI} the
value is
$10^{-6}$ s). Thus, one may suggest that the elastic motion is
following the proton tunnelling without inertia so that
we may take into account the lattice deformation
only through the potential energy $H_Y$, neglecting the kinetic one.
In writing $H_Y$ we shall follow the method worked out in papers
\cite{GKY}, \cite{PEY}, aiming at a simplified description of the
dynamics of the double helix which is  considered  as a
one-dimensional lattice of vectors ${\bf y}_n$ that describe the mutual
positions of the two strands at sites corresponding to the base pair of
index $n$. The helix structure is described with the help of the
covariant derivative for the description of deformations resulting from
the change in the positions of the strands, so that the potential $H_Y$ reads 
\begin{eqnarray}
\label{a12}
H_Y =  \sum _{n}\left [\frac{K}{2}(\nabla {\bf y}_n)^2
+ \frac{\epsilon }{2}{\bf y_n}^2 \right ] \, . 
\end{eqnarray}
Here the harmonic term $(1/2) \epsilon y_n^2$ describes the binding
of the two strands, and $\nabla {\bf y}_n$ is the covariant derivative 
in discrete form,
linearized with respect to rotations that relates the two adjacent sites,
$n$ and $n+1$.
\begin{eqnarray}
\label{a2}
\nabla {\bf y}_n \equiv \frac{1}{a}({\bf y}_{n+1} - {\bf y}_n + {\hat
\Omega }{\bf y}_n) \, , 
\end{eqnarray}
and
\begin{eqnarray} 
\label{a3}\Omega _{ij} = - \epsilon _{ijk} \Omega _k
\end{eqnarray}
where $\Omega _k \equiv (0 \, , 0\, , \Omega )$, $K \, , \epsilon $ are elastic
constants, and $a$ is the spacing of the lattice of the base pairs.
We are assuming that the molecule is parallel to the $z$ axis.
Having chosen the dynamical variables $b_n^+\, , b_n \, , {\bf y}_n$ for the protons and the elastic
excitations, we have a limited number of option for the interaction energy. The simplest one should couple
the proton excited states given by $b_n^+b_n$, and the elastic deformations, which are a function of ${\bf
y}_n$, in the simplest situation we adopt, the linear one. We assume that ${\bf y}_n$ enters in the form
of its covariant derivative, for this way one might accommodate the interaction of the proton tunnelling
with the stack of the $\pi -$electrons of the bases. 

The choice of the interaction Hamiltonian is delicate but the
knowledge of the structure and binding in DNA gives some indications.
The stacking interaction between adjacent
base pairs is strongly affected by the overlap of the $\pi $ electrons
of the bases. But the proton transfer is accompanied by a
redistribution of the electrons on the bases so that the stacking
interaction is changed and therefore we expect that the proton
tunneling will affect the {\em coupling} between adjacent bases. It
means that the derivative $\nabla {\bf y}_n$ rather than ${\bf y}_n$ itself
should enter in the interaction Hamiltonian. We have chosen the
simplest form coupling the commponent of the gradient which is along
the direction ${\bf h}_n$
of the hydrogen bonds connecting the bases within pair $n$ and
the state of the protons in this pair given by $b_n^+b_n$, i.e.
\begin{eqnarray}
H_I = - \lambda \sum _n(\nabla {\bf y}_n \cdot {\bf h}_n)b_n^+b_n \, ,
\label{g1}
\end{eqnarray}
in which the vectors ${\bf h}_n$ are subject to the helical symmetry
of the molecule. They can be written in the form
\begin{eqnarray}
\label{a4}
{\bf h}_n = (\cos n\alpha \, ; \sin n\alpha \, ; 0) \, ,
\end{eqnarray}
where the angle $\alpha
$ is related to the helical pitch. 
Having chosen a primitive model of the DNA, and 
neglecting the subtle features like 2 or 3 hydrogen bonds 
for the A -T or G - C base pairs, and confining
ourselves to one-site one H-bond picture, we may set $\alpha = \Omega$ 
equalizing the pitch of the ${\bf
h}_n$ and the twist of the double helix.

It is important to notice that the sign of $\lambda $ will be crucial 
for the SNLNL equation that we shall derive.
Our choice means that the stacking energy is 
reduced when protons are in the excited state.

We employ the method worked out by Davydov \cite{DA} 
to study the model described above. According to
\cite{DA},
\begin{itemize}
\item we have to calculate the effective potential
\begin{eqnarray}
\label{b1}
U_{eff} = <D(t) |H_Y + H_I|D(t) >
\end{eqnarray}
which is a function of ${\bf y}_n$, 

\item find its minimum, ${\bf y}^0_n$, 

\item substitute ${\bf y}^0_n$ into $H$
given by Eq. (\ref{a1}), thus obtaining the effective Davydov Hamiltonian $H_D$. 
\end{itemize}

Finally, we have to
write down the Schrodinger equation for the function $|D(t)>$, given by Eq. (\ref{b111}), and the
Hamiltonian $H_D$
\begin{eqnarray}
\label{g2}
i\hbar \frac{\partial }{\partial t}|D(t) = H_D |D(t)>
\end{eqnarray}
Both sides of the equation indicated above are linear forms in the operators $b_n^+$, and by equating the
coefficients at corresponding $b_n^+$, we obtain the following equation for $A_n$
\begin{eqnarray}
\label{b8}
i \hbar \frac{\partial A_n}{\partial t}= E_0 A_n - \kappa (A_{n+1} + A_{n-1}) -
\frac{\lambda ^2}{K}|A_n|^2 A_n - \frac{\lambda ^2}{K}[\sum _m |A_m|^4]A_n +
\end{eqnarray}
\begin{eqnarray}
\nonumber\frac{\lambda ^2}{2 K}\frac{\epsilon a^2}{K\Omega ^2}\{\sum _{m^\prime , m^{\prime \prime }}
\cos^{|m^\prime - m^{\prime \prime }|}\phi 
\cos [(m^\prime - m^{\prime \prime })(\phi - \Omega )]
|A_{m^\prime }|^2|A_{m^{\prime \prime }}|^2\}A_n+
\end{eqnarray}
\begin{eqnarray}
\nonumber
\frac{\lambda ^2}{2 K}\frac{\epsilon a^2}{K\Omega ^2}
\{\sum _m\cos^{|m-n|}\phi \cos [(m-n)(\phi - \Omega )]|A_m|^2\}A_n \, ,
\end{eqnarray}
in which the angle $\phi $ is determined by the equation $\tan  
\phi = \Omega $ 
and $\lambda /K$ and $\epsilon a^2/(K \Omega ^2) \ll 1$ are small parameters.

In assessing the importance of $\cos $- terms in the equation
given above, it is worthwhile to note that since we have assumed
$\alpha = \Omega $ and there is the relation $\tan  \phi = \Omega $, 
the typical term
in the equation (\ref{b8}) contains a factor that reads asymptotically
$$
\exp(-\Omega ^2 |m - n|) cos(|m-n|\Omega ^3)
$$
We have assumed $\Omega $ being a small parameter, and therefore the
oscillations due to $\Omega ^3$ are negligible.
Let us consider a possible simplification of the equation (\ref{b8}) 
obtained above.
There is a chance that it may have a bearing on the dynamics of the 
proton transfer in DNA.
To that end note that the angles $\phi $ and $\Omega $ 
are close to each other; indeed, for the B DNA the
angle $\Omega $ corresponds to the pitch, i.e. 10 steps for $2\pi $. 
For this reason, we set all the
functions $\cos[(m-n)(\phi - \Omega )]$ equal to 1 in (\ref{b8}), and 
introduce
\begin{eqnarray}
\label{g3}
A(z , t) = \exp (-i \frac{E_0 t}{\hbar }) B(z , t)
\end{eqnarray}
Then, in the continues notations, 
(i.e. for scales larger than $a$) 
we obtain the following non-linear and non-local Schrodinger equation ($z \equiv x/a$).
\begin{eqnarray}
\label{b9}
i \frac{\partial B}{\partial t}= - \omega _H \frac{\partial ^2 B}{\partial z^2}
-\frac{\lambda }{K}\omega _T|B^2|B - \frac{\lambda }{K} \omega _T B\int |B(z^\prime )|^4 d z^\prime 
+
\end{eqnarray}
\begin{eqnarray}
\nonumber\frac{\lambda }{K}\omega _T \frac{\epsilon a^2}{K\Omega ^2}(\int dz^\prime \int dz^{\prime \prime }
\exp(-\mu |z^\prime - z^{\prime \prime }|) |B(z^\prime )|^2|B(z^{\prime \prime }|^2) B(z) 
+
\end{eqnarray}
\begin{eqnarray}
\nonumber
\frac{\lambda }{K}\omega _T \frac{\epsilon a^2}{K\Omega ^2}(\int dz^\prime 
\exp(-\mu |z-z^\prime |) |B(z^\prime |^2) B(z)\, ,
\end{eqnarray}
in which
\begin{eqnarray}
\label{b10}\mu = - \ln\cos \phi = \frac{1}{2} \ln (1 + \Omega ^2) \simeq \frac{\Omega ^2}{2} \, ,
\end{eqnarray}
and $\omega _H \equiv \kappa /\hbar $, $\omega _T \equiv \lambda /\hbar $.
We may perform a very crude estimate so as to see the part played by the non-local terms as regards the
structure of solitons that might turn around.

The standard procedure to treat the non-linear Schrodinger equation is as follows 
\cite{DA}, \cite{scot}.
We are looking for the solution in the propagating wave ({\it soliton}) form
\begin{eqnarray}
\label{g4}
B = \exp[i(kz-\omega t)]\psi (z- v t)
\end{eqnarray}
Introducing (\ref{g4}) into (\ref{b9}) we get two equations for the imaginary and reel parts of the
solution (\ref{g4}).
{From} the former one we find the velocity of the soliton $v$
\begin{eqnarray}
\label{g5}
v = 2 \omega _H k \, ,
\end{eqnarray}
and the real part of (\ref{b9}) leads to the equation
\begin{eqnarray}
\label{g6}
\omega \psi = - \omega _H[-k^2 + {\dot {\dot \psi }}] - \frac{\lambda }{K}\omega _T\psi ^3
-\frac{\lambda }{K}\omega _T \psi \int |\psi |^4 dz^\prime + 
\end{eqnarray}
\begin{eqnarray}
\nonumber
\frac{\lambda }{K}\omega _T \frac{\epsilon
a^2}{K \Omega ^2} \psi \int dz^\prime \int dz^{\prime \prime } 
\exp(-\mu |z^\prime - z^{\prime \prime
})|\psi (z^\prime )|^2|\psi (z^{\prime \prime })|^2 + 
\end{eqnarray}
\begin{eqnarray}
\nonumber
\frac{\lambda }{K} \omega _T \psi \int dz^{\prime }
\exp(-\mu |z - z^\prime |)|\psi (z^\prime )|^2
\end{eqnarray}
Neglecting $\mu $ we get the standard non-linear Schrodinger equation however with renormalized
coefficients. It is easy to find the first integral of the equation, i.e. the energy $W$
\begin{eqnarray}
\label{g7}
(\dot \psi )^2 + [-k^2 + \nu - \frac{\lambda }{K}\xi (\frac{\epsilon a^2}{K\Omega ^2} C + \Gamma )]\psi ^2 +
\frac{\lambda }{2 K}\xi \psi ^4 = W
\end{eqnarray}
Where we used the following notations:
$$
\xi = \frac{\omega _T}{\omega _H} \, , \, \nu = \frac{\omega }{\omega _H} \, , 
\, \Gamma = \int dz^\prime |\psi
(z^\prime )|^4
$$
and $C$ is a factor of the order of 1.
It is worth noting that even though we have made rough simplifications,
as regards the non-locality, its bearing on the proton dynamics
still has remained, as is seen in the $\epsilon a^2/K\Omega ^2 $ term, preserved
in the equation given above. One may infer from the fact that under
approximations used, there is a profound interplay of the dynamics of proton transfer and the
conformational structure of the double helix.

{From} the equation (\ref{g6}) we obtain the asymptotic 
width $\Delta $ of the soliton
which reads
\begin{eqnarray}
\label{end}
\Delta = \left [\frac{\omega _T^2}{\omega _{ac} 
\omega _H}\frac{\epsilon a^2}{K\Omega ^2}\right ]^{-1/2} \, .
\end{eqnarray}
The frequency $\omega _{ac} \equiv K/\hbar $ is generally accepted to be within
$10^{11} - 10^{13} $Hz, but the available estimates for $\omega _T \, , \omega _H \, ,
\epsilon $ widely diverge, $\omega _H = 10^6 - 10^{11}$ Hz, $\omega _T/\omega _H =
0.1 - 0.01$, and $\epsilon a^2/K\Omega ^2 \simeq 0.1$. Consequently, the width
of the solution varies within $10 - 1000 \AA $, so that one may expect
a concerted tunnelling of protons for the lower estimation of $\Delta $.

III.

Concluding, we should like to draw attention to the fact that the
existence of an appreciable interaction between the proton transfer
inside the hydrogen bonds of the double helix, and elastic modes of
the latter, could result in a concerted dynamics of the protons which
is generally of non-linear character and governed by the non-linear
and non-local Schrodinger equation. To our knowledge, it is for the
first time one is running across the non-linear Schrodinger equation
with non-local terms. The transfer of protons may be due to various
reasons, and among these are the action of external agents,
especially, according to recent experimental work \cite{?} , \cite{??}
enzymes; it is also worth keeping in mind the possible links with the
mutation mechanism, \cite{SA}.
In this paper we have tried to put these ideas in more quantitative
form within the framework of a model that accommodates the basic
symmetry structure of the double helix, and as was shown above, allows
for certain rough estimates of dynamical features that may surface;
presumably of the soliton nature. In this respect, it is worthwhile to
note that even the crude estimate we made, conserves the bearing of
the double helix, as in seen through the occurrence of the
characteristic parameter $\epsilon /(K \Omega ^2)$ in
the final formulas.  Of course the equation (\ref{b8}) we obtained has
a larger scope. It may not imply the existence of
solitons in DNA, but could open another possibility, the possible
existence of nonlinear localization leading to a collapse of an
initially broad excitation into a higly localized deformation
\cite{ZA}. Collapse does not occur in the
standard NLS equation, but the existence of nonlinearities with higher
power in our NLNLS suggest that it could occur in this equation,
although, as these terms are non local, no definite conclusion can be
given without further studies of the equation. If this hypothesis
would be confirmed, a weak and broad perturbation of the hydrogen
bonds of a DNA molecule by the vicinity of an enzyme carrying local
charges could trigger this nonlinear localization phenomenon and
finally lead to the formation of a tutomerized form by the tunneling
of one proton. As the present stage of the study, this is however only
a speculation raised by the form of the NLNLS equation that we derived.

\ \ \acknowledgements

The research described in this publication was made possible in part by 
RFFR Grant 00-02-17785.
 One of the authors (V.L.G.) is thankful to the Lab. Phys. ENS Lyon,
for the hospitality, and the programme of 
exchange between ENS and Landau Institute for Theoretical
Physics for the opportunity to participate in the programme. 
Fruitful discussions with E.Kuznetsov and Yu.M. Yevdokimov,
and the useful correspondence with J.L.Leroy and F.Fillaux are 
gratefully acknowledged.

\end{document}